\documentclass[12pt]{article}
\usepackage{graphicx}
\newcommand{\bers}{\begin{eqnarray*}}
\newcommand{\eers}{\end{eqnarray*}}
\newcommand{\bt}{\begin{itemize}}
\newcommand{\et}{\end{itemize}}
\def\beq{\begin{equation}}
\def\eeq{\end{equation}}
\def\bea{\begin{eqnarray}}
\def\eea{\end{eqnarray}}
\def\nn{\nonumber}
\def\sss{\scriptscriptstyle}
\def\bd{B^0}
\def\bdbar{{\overline{B^0}}}

\def\barp{{\raise.35ex\hbox
{${\sss (}$}}---{\raise.35ex\hbox{${\sss )}$}}}
\def\bdbarp{\hbox{$B_d$\kern-1.4em\raise1.4ex\hbox{\barp}}}
\def\bsbarp{\hbox{$B_s$\kern-1.4em\raise1.4ex\hbox{\barp}}}

\def\roughly#1{\mathrel{\raise.3ex\hbox
{$#1$\kern-.75em\lower1ex\hbox{$\sim$}}}}

\def\adirpm{{a_{\sss dir}^{+-}}}
\def\adir00{{a_{\sss dir}^{00}}}
\def\alphaeff{{\alpha_{\sss eff}}}
\def\epjc#1#2#3{{\it Eur.\ Phys.\ J.}\ {\bf C#1} (#2) #3}

\def\plb#1#2#3{{\it Phys.\ Lett.} {\bf B#1} (#2) #3}
\def\prd#1#2#3{{\it Phys.\ Rev.} {\bf D#1} (#2) #3}
\def\newprd#1#2#3{{\it Phys.\ Rev.} {\bf D#1} (#2) #3}

\def\prl#1#2#3{{\it Phys.\ Rev.\ Lett.} {\bf #1} (#2) #3}

\newread\epsffilein 
\newif\ifepsffileok 
\newif\ifepsfbbfound 
\newif\ifepsfverbose 
\newdimen\epsfxsize 
\newdimen\epsfysize 
\newdimen\epsftsize 
\newdimen\epsfrsize 
\newdimen\epsftmp 
\newdimen\pspoints 
\def\epsfbox#1{\global\def\epsfllx{72}\global\def\epsflly{72}%
 \global\def\epsfurx{540}\global\def\epsfury{720}%
 \def\lbracket{[}\def\testit{#1}\ifx\testit\lbracket
 \let\next=\epsfgetlitbb\else\let\next=\epsfnormal\fi\next{#1}}%
\def\epsfgetlitbb#1#2 #3 #4 #5]#6{\epsfgrab #2 #3 #4 #5 .\\%
 \epsfsetgraph{#6}}%
\def\epsfnormal#1{\epsfgetbb{#1}\epsfsetgraph{#1}}%
\def\epsfgetbb#1{%
\openin\epsffilein=#1
\ifeof\epsffilein\errmessage{I couldn't open #1, will ignore it}\else
 {\epsffileoktrue \chardef\other=12
 \def\do##1{\catcode`##1=\other}\dospecials \catcode`\ =10
 \loop
 \read\epsffilein to \epsffileline
 \ifeof\epsffilein\epsffileokfalse\else
 \expandafter\epsfaux\epsffileline:. \\%
 \fi
 \ifepsffileok\repeat
 \ifepsfbbfound\else
 \ifepsfverbose\message{No bounding box comment in #1; using defaults}\fi\fi
 }\closein\epsffilein\fi}%
\def\epsfclipstring{}
\def\epsfsetgraph#1{%
 \epsfrsize=\epsfury\pspoints
 \advance\epsfrsize by-\epsflly\pspoints
 \epsftsize=\epsfurx\pspoints
 \advance\epsftsize by-\epsfllx\pspoints
 \epsfxsize\epsfsize\epsftsize\epsfrsize
 \ifnum\epsfxsize=0 \ifnum\epsfysize=0
 \epsfxsize=\epsftsize \epsfysize=\epsfrsize
 \epsfrsize=0pt
 \else\epsftmp=\epsftsize \divide\epsftmp\epsfrsize
 \epsfxsize=\epsfysize \multiply\epsfxsize\epsftmp
 \multiply\epsftmp\epsfrsize \advance\epsftsize-\epsftmp
 \epsftmp=\epsfysize
 \loop \advance\epsftsize\epsftsize \divide\epsftmp 2
 \ifnum\epsftmp>0
 \ifnum\epsftsize<\epsfrsize\else
 \advance\epsftsize-\epsfrsize \advance\epsfxsize\epsftmp \fi
 \repeat
 \epsfrsize=0pt
 \fi
 \else \ifnum\epsfysize=0
 \epsftmp=\epsfrsize \divide\epsftmp\epsftsize
 \epsfysize=\epsfxsize \multiply\epsfysize\epsftmp
 \multiply\epsftmp\epsftsize \advance\epsfrsize-\epsftmp
 \epsftmp=\epsfxsize
 \loop \advance\epsfrsize\epsfrsize \divide\epsftmp 2
 \ifnum\epsftmp>0
 \ifnum\epsfrsize<\epsftsize\else
 \advance\epsfrsize-\epsftsize \advance\epsfysize\epsftmp \fi
 \repeat
 \epsfrsize=0pt
 \else
 \epsfrsize=\epsfysize
 \fi
 \fi
 \ifepsfverbose\message{#1: width=\the\epsfxsize, height=\the\epsfysize}\fi
 \epsftmp=10\epsfxsize \divide\epsftmp\pspoints
 \vbox to\epsfysize{\vfil\hbox to\epsfxsize{%
 \ifnum\epsfrsize=0\relax
 \includegraphics{#1}%
 \else
 \epsfrsize=10\epsfysize \divide\epsfrsize\pspoints
 \includegraphics{#1}%
 \fi
 \hfil}}%
\global\epsfxsize=0pt\global\epsfysize=0pt}%
 {\catcode`\%=12 \global\let\epsfpercent=
\long\def\epsfaux#1#2:#3\\{\ifx#1\epsfpercent
 \def\testit{#2}\ifx\testit\epsfbblit
 \epsfgrab #3 . . . \\%
 \epsffileokfalse
 \global\epsfbbfoundtrue
 \fi\else\ifx#1\par\else\epsffileokfalse\fi\fi}%
\def\epsfempty{}%
\def\epsfgrab #1 #2 #3 #4 #5\\{%
\global\def\epsfllx{#1}\ifx\epsfllx\epsfempty
 \epsfgrab #2 #3 #4 #5 .\\\else
 \global\def\epsflly{#2}%
 \global\def\epsfurx{#3}\global\def\epsfury{#4}\fi}%
\def\epsfsize#1#2{\epsfxsize}
\pagestyle{plain}

\begin{document}

\begin{flushright}
TECHNION-PH-2001-27\\
UdeM-GPP-TH-01-89\\
IMSc-2001/05/27 \\
\end{flushright}

\begin{center}

{\large \bf
\centerline{Improving Bounds on Penguin Pollution in $B \to \pi\pi$}}
\vspace*{1.0cm}
{\large
Michael Gronau$^{a,}$\footnote{gronau@physics.technion.ac.il},
David London$^{b,}$\footnote{london@lps.umontreal.ca},
Nita Sinha$^{c,}$\footnote{nita@imsc.ernet.in}\\
and Rahul Sinha$^{c,}$\footnote{sinha@imsc.ernet.in}}
\vskip0.3cm
{\it ${}^a$ Department of Physics, Technion-Israel Institute of Technology,}\\
{\it Technion City, 32000 Haifa, Israel}\\
\vskip0.3cm
{\it ${}^b$ Laboratoire Ren\'e J.-A. L\'evesque, Universit\'e de
  Montr\'eal,} \\
{\it C.P. 6128, succ.\ centre-ville, Montr\'eal, QC, Canada H3C 3J7} \\
\vskip0.3cm
{\it ${}^c$ Institute of Mathematical Sciences, Taramani,
 Chennai 600113, India}\\
\vskip0.5cm
\bigskip
(\today)
\vskip0.5cm
{\Large Abstract\\}
\vskip3truemm
\parbox[t]{\textwidth} {In the presence of penguin contributions, the
  indirect CP asymmetry in $\bd(t)\to\pi^+\pi^-$ measures $\sin
  (2\alpha + 2\theta)$, where $2\theta$ parametrizes the size of the
  penguin ``pollution.'' We derive a new upper bound on $|2\theta|$,
  requiring the measurement of $BR(B^+\to\pi^+\pi^0)$ and an upper
  bound on $B^{00} \equiv {1\over 2} [BR(\bd \to \pi^0\pi^0) +
  BR(\bdbar \to \pi^0\pi^0)]$. The new bound is stronger than those
  previously discussed in the literature.  We also present a lower
  bound on $B^{00}$. Current data may suggest that it is not very
  small, in which case $\theta$ can be determined using a complete
  isospin analysis.}
\end{center}
\thispagestyle{empty}
\newpage
\setcounter{page}{1}
\textheight 23.0 true cm
\baselineskip=14pt

Over the past decade or so, a great deal of attention has been
focussed on CP violation in the $B$ system. By measuring $\alpha$,
$\beta$ and $\gamma$, the three interior angles of the unitarity
triangle, it will be possible to test the standard model (SM)
explanation of CP violation \cite{CPreview}. Indeed, the first
measurements of $\beta$ have already been reported \cite{betameas},
and it is hoped that we will soon have definitive evidence of CP
violation in $B$ decays.

For the measurement of the angle $\alpha$, a principal decay mode
considered is $\bd(t) \to \pi^+ \pi^-$. (The decays
  $\bd(t)\to\rho\pi\to\pi^+\pi^-\pi^0$ \cite{Dalitz} and $B^0_{d,s}(t)
  \to K^{(*)} {\bar K}^{(*)}$ \cite{BKK} can also be used to cleanly
  obtain $\alpha$.) Unfortunately, this mode suffers from a
well-known problem: penguin contributions may be large
\cite{penguins}, and their presence will spoil the clean extraction of
$\alpha$. This problem of penguin ``pollution'' can be eliminated with
the help of an isospin analysis \cite{isospin}. By measuring the rates
for $B^+ \to \pi^+ \pi^0$ and $\bd/\bdbar \to \pi^0\pi^0$, in addition
to $\bd(t)\to\pi^+\pi^-$, the penguin contributions can be eliminated
so that $\alpha$ can again be measured cleanly.

However, the isospin analysis itself suffers from a potential
practical complication: it requires separate measurements of $BR(\bd
\to \pi^0\pi^0)$ and $BR(\bdbar \to \pi^0\pi^0)$. This may be a
problem for several reasons. First, these branching ratios are
expected to be smaller than $\bd\to\pi^+\pi^-$. Second, the presence
of two $\pi^0$'s in the final state means that the reconstruction
efficiency is also smaller. And third, in order to measure the two
branching ratios individually, it will be necessary to tag the
decaying $\bd$ or $\bdbar$ meson, which will further reduce the
measurement efficiency. The upshot is that it may not be possible to
measure either of these two branching ratios, or we may only have
information (i.e.\ an actual measurement or an upper limit) on the sum
of the branching ratios. In either case, a full isospin analysis
cannot be carried out.

But this then begs the question: assuming that we have, at best, only
partial knowledge of the sum of $BR(\bd \to \pi^0\pi^0)$ and
$BR(\bdbar \to \pi^0\pi^0)$, can we at least put bounds on the size of
penguin pollution? To be more precise: in the presence of penguin
amplitudes, the CP asymmetry in $\bd(t)\to\pi^+\pi^-$ does not measure
$\sin 2\alpha$, but rather $\sin (2\alpha + 2\theta)$, where $2\theta$
parametrizes the effect of the penguin contributions.  Is it
possible to constrain $\theta$?  As demonstrated by Grossman and Quinn
\cite{GQ}, the answer to this question is yes.  They were able to show
that $|2\theta|$ can be bounded even if we have only an upper limit on
the sum of $BR(\bd \to \pi^0\pi^0)$ and $BR(\bdbar \to \pi^0\pi^0)$.
Charles \cite{Charles} also examined this question, and found an
improvement to the Grossman-Quinn bound involving the direct asymmetry
in $B^0 \to \pi^+\pi^-$, as well as an independent bound involving
different measurements.

The main purpose of this Letter is to present a new bound on
$|2\theta|$ which is an improvement on both the Grossman-Quinn and
Charles bounds.  In contrast to the earlier bounds, the new bound
follows from the requirements that the two isospin triangles close and
have a common base, making it the most stringent bound possible on
$|2\theta|$.  Indeed, the new bound contains the two previous bounds
as limiting cases. We also present the constraints on the sum of
$BR(\bd \to \pi^0\pi^0)$ and $BR(\bdbar \to \pi^0\pi^0)$ which follow
from the requirement of closure of the triangles. As we will show, if
$BR(B^+ \to \pi^+\pi^0)/BR(B^0 \to \pi^+\pi^-)$ is larger than one, as
present experimental central values suggest, the branching ratios for
$\bd/\bdbar\to\pi^0\pi^0$ cannot be tiny. In this case, it may well be
possible to carry out the full isospin analysis.  Finally, we show how
measurements of $\bd(t)\to\pi^+\pi^-$ alone can be used to place a
lower limit on the magnitude of the penguin amplitude.

We begin with a brief review of the bounds of Grossman-Quinn and
Charles. Defining
\bea
B^{+-} & \equiv & {1\over 2} \left( |A^{+-}|^2+|{\bar A^{+-}}|^2 \right) ~,
\nn\\
\adirpm & \equiv & {{|A^{+-}|^2-|{\bar A^{+-}}|^2} \over
{|A^{+-}|^2+|{\bar A^{+-}}|^2} } ~, \nn\\
B^{00} & \equiv & {1\over 2} \left( |A^{00}|^2+|{\bar A^{00}}|^2 \right) ~,
\nn\\
B^{+0} & \equiv & |A^{+0}|^2 ~,
\label{observables}
\eea
where $A^{+-}$ and ${\bar A^{+-}}$ are the amplitudes for $\bd \to
\pi^+\pi^-$ and $\bdbar \to \pi^+\pi^-$, respectively, and similarly
for $A^{00}$, ${\bar A^{00}}$ and $A^{+0}$, Grossman and Quinn
obtained the bound \cite{GQ}
\beq
\cos 2\theta \ge 1 - 2 {B^{00} \over B^{+0}} ~.
\label{GQbound1}
\eeq
In Ref.~\cite{Charles}, Charles noted that this bound can be improved:
\beq
\cos 2\theta \ge {1 - 2 B^{00} / B^{+0} \over y} ~,
\label{GQbound}
\eeq
where $y \equiv\sqrt{1-(\adirpm)^2}$. In what follows, we will refer
to this as the Grossman-Quinn bound. Charles also pointed out the
existence of a second bound:
\beq
\cos 2\theta \ge {1 - 4 B^{00} / B^{+-} \over y} ~,
\label{Charlesbound}
\eeq
involving a different ratio of rates.  This will be referred to as the
Charles bound. From either Eq.~(\ref{GQbound}) or
(\ref{Charlesbound}), one sees that, given a measurement of $B^{00}$,
one may be able to put a nontrivial lower bound on $\cos 2\theta$,
i.e.\ an upper bound on $|2\theta|$. Even if one has only an upper bound
on $B^{00}$, this still yields a lower limit on $\cos 2\theta$. Thus,
partial information about $BR(\bd \to \pi^0\pi^0)$ and $BR(\bdbar \to
\pi^0\pi^0)$ does indeed allow us to constrain the size of penguin
pollution. We note, however, that neither Eq.~(\ref{GQbound}) nor
(\ref{Charlesbound}) involves all three charge-averaged decay rates,
$B^{+-}$, $B^{+0}$ and $B^{00}$. Thus, a condition for the closure of
the two isospin triangles is not included in these bounds.

We now turn to our new bound on $|2\theta|$, which is the strongest
possible bound on this quantity. We assume that the charge-averaged
rates $B^{+-}$ and $B^{+0}$ have been measured, and that we have (at
least) an upper bound on $B^{00}$. We will present in detail a
geometrical derivation. A second algebraic proof, which gives this
bound more directly, will also be outlined.

In the presence of penguin contributions, the $B\to\pi\pi$ decay
amplitudes take the form
\bea
{1 \over \sqrt{2}} A^{+-} & = & T e^{i\gamma} + P e^{-i\beta} ~, \nn\\
A^{00} & = & C e^{i\gamma} - P e^{-i\beta} ~, \nn\\
A^{+0} & = & (C + T) e^{i\gamma} ~,
\label{amps}
\eea
where the complex amplitudes $T$, $C$ and $P$, which are sometimes
referred to as ``tree", ``colour-suppressed" and ``penguin"
amplitudes, include strong phases. Note that we have implicitly
imposed the isospin triangle relation
\beq
{1 \over \sqrt{2}} A^{+-} + A^{00} = A^{+0} ~.
\eeq
The ${\bar A}$ amplitudes can be obtained from the $A$ amplitudes by
simply changing the signs of the weak phases.

\begin{figure}
\centerline{\epsfxsize 5.0 truein \epsfbox {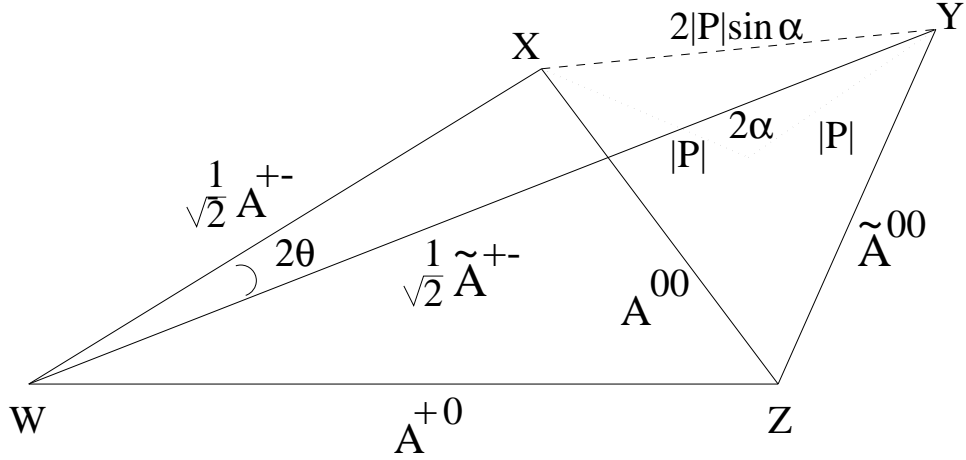}}
\caption{The $A$ and ${\tilde A}$ isospin triangles.}
\label{isotriangles}
\end{figure}

It is convenient to define the new amplitudes ${\tilde A}^{ij} \equiv
e^{2 i \gamma} {\bar A}^{ij}$. Then three observations can be made.
First, ${\tilde A}^{-0} = A^{+0}$, so that the $A$ and ${\tilde A}$
triangles have a common base. (A tiny electroweak penguin amplitude,
forming a very small angle between $A^{+0}$ and ${\tilde A}^{-0}$, can
be taken into account analytically \cite{GPY}. However, here it will
be neglected.) Second, in the absence of penguin contributions,
${\tilde A}^{+-} = A^{+-}$. Thus, the relative phase $2\theta$ between
these two amplitudes is due to penguin pollution.  Third, the relative
phase between the penguin contributions in ${\tilde A}^{00}$ and
$A^{00}$ is $2(\beta + \gamma) \sim 2\alpha$. All this information is
encoded in Fig.~\ref{isotriangles}. Note that the distance between the
points $X$ and $Y$ is $2 \ell \equiv 2 |P| \sin\alpha$.

Now, $|P|$ can be expressed in terms of observables \cite{Charles}:
\beq
|P|^2 = { B^{+-} \over 4 \sin^2 (\alphaeff - \theta)} \left[1-y\cos(2
  \alphaeff - 2 \alpha)\right] ~,
\label{P+-}
\eeq
where $2\alphaeff=2\alpha + 2\theta$ is the relative phase between the
$A^{+-}$ and $e^{-2i\beta}{\bar A}^{+-}$ amplitudes, occurring in the
time-dependent rate of $\bd(t)\to\pi^+\pi^-$,
\beq
\Gamma(B^0(t)\to\pi^+\pi^-) = e^{-\Gamma t}B^{+-}\left [1 + \adirpm
\cos(\Delta mt) - y\sin 2\alphaeff\sin(\Delta mt)\right ]~.
\label{B0(t)}
\eeq
We therefore can write
\beq
\ell = {1 \over 2} \sqrt{B^{+-}} \sqrt{1 - y \cos 2\theta} ~.
\label{elldef}
\eeq
Thus, a constraint on $\ell$ implies a bound on $\cos 2\theta$.

In order to constrain $\ell$, we proceed as follows. First, we assign
a coordinate system to Fig.~\ref{isotriangles} such that the origin is
at the midpoint of the points $X$ and $Y$. The points $X$ and $Y$
correspond respectively to the coordinates $(+\ell, 0)$ and $(-\ell,
0)$. The points $W$ and $Z$ are labelled respectively as $(x_1, y_1)$
and $(x_2, y_2)$. The goal of the exercise is to find the values of
these four coordinates. We then note that
\bea
{1\over 2} \left| A^{+-} \right|^2 & = & (x_1 - \ell)^2 + y_1^2 ~, \nn\\
{1\over 2} \left| {\tilde A}^{+-} \right|^2 & = & (x_1 + \ell)^2 + y_1^2 ~.
\eea
Assuming $B^{+-}$ and $a_{dir}^{+-}$ have been measured, we can solve
for $x_1$ and $y_1$ (up to a discrete ambiguity) as a function of
$\ell$. We also note that
\bea
\left| A^{00} \right|^2 & = & (x_2 - \ell)^2 + y_2^2 ~, \nn\\
\left| {\tilde A}^{00} \right|^2 & = & (x_2 + \ell)^2 + y_2^2 ~.
\eea
Here we assume that only information about $B^{00}$ is available, so
this gives us only one equation in the two unknowns $x_2$ and $y_2$.
However, we also have
\beq
\left| A^{+0} \right|^2 = (x_1 - x_2)^2 + (y_1 - y_2)^2 ~,
\eeq
So this gives us another equation involving $x_2$ and $y_2$. We
therefore have four (nonlinear) equations in four unknowns, and we can
solve for these coordinates as a function of $\ell$. The equations
are:
\bea
B^{+-} & = & 2 (x_1^2 + y_1^2) + 2 \ell^2 ~,\nn\\
B^{+-} a_{dir}^{+-} & = & - 4 x_1 \ell ~,\nn\\
B^{00} & = & (x_2^2 + y_2^2) + \ell^2 ~,\nn\\
B^{+0} & = & (x_1^2 + y_1^2) + (x_2^2 + y_2^2) - 2 x_1 x_2 - 2 y_1 y_2 ~.
\eea

However, the key point is the following: we must obtain only real
solutions for $x_2$ and $y_2$, otherwise the triangles do not close.
This puts a constraint on $\ell$, which, with a bit of simple algebra,
can be written
\beq
x_1^2 C_1^2 - (x_1^2 + y_1^2) (C_1^2 - C_2 y_1^2) \ge 0 ~,
\eeq
where
\bea
C_1 & \equiv & {1 \over 2} \left( {1\over 2} B^{+-} - B^{+0} +  B^{00}
  -  2\ell^2 \right)~,\nn\\
C_2 & \equiv & B^{00} - \ell^2 ~.
\eea
This leads to
\beq
\ell^2 \le { 2 B^{+-} B^{00} - \left( {1\over 2} B^{+-} - B^{+0} +
B^{00}\right)^2 \over 4 B^{+0} } ~,
\eeq
or, using the expression for $\ell$ in Eq.~(\ref{elldef}),
\beq
\cos 2\theta \ge { \left( {1\over 2}B^{+-} + B^{+0} - B^{00} \right)^2 -
  B^{+-} B^{+0} \over B^{+-} B^{+0} y} ~.
\label{superbound}
\eeq
This is the new lower bound on $\cos 2\theta$ (or upper bound on
$|2\theta|$). We reiterate that this bound has been derived assuming
that the isospin triangles close and have a common base. Thus, to the
extent that isospin is violated, whether by electroweak penguin
contributions or by $\pi^0$--$\eta,\eta'$ mixing \cite{Gardner}, the
bound will be correspondingly weakened.

Note that this lower bound on $\cos 2\theta$ can be written
\beq
\cos 2\theta \ge {1 - 2 B^{00} / B^{+0} \over y} + { \left( B^{+-} - 2
    B^{+0} + 2 B^{00} \right)^2 \over 4 B^{+-} B^{+0} y} ~.
\label{SBGQequiv}
\eeq
The first term is simply the Grossman-Quinn bound of
Eq.~(\ref{GQbound}). Since the second term is always positive, the new
bound is stronger than the Grossman-Quinn bound.  Similarly,
Eq.~(\ref{superbound}) can be written
\beq
\cos 2\theta \ge {1 - 4 B^{00} / B^{+-} \over y} + { \left( B^{+-} - 2
    B^{+0} - 2 B^{00} \right)^2 \over 4 B^{+-} B^{+0} y} ~,
\eeq
where the first term is the Charles bound of Eq.~(\ref{Charlesbound}).
The second term is positive, so that once again the new bound is more
constraining than the Charles bound. Note that neither of the previous
bounds fully uses the requirement that the isospin triangles close. By
contrast, all isospin information has been used in obtaining
Eq.~(\ref{superbound}), so that this is the most stringent possible
bound on $\cos 2\theta$.

An alternative way of deriving the new bound of Eq.~(\ref{superbound})
involves the direct calculation of the minimum of $\cos 2\theta$ under
the assumption that $B^{+-}$, $B^{+0}$, $B^{00}$ and $y$ are given.
Let us define $\Phi$ to be the angle between $A^{+-}$ and $A^{+0}$,
and $\bar\Phi$ to be the angle between ${\tilde A}^{+-}$ and ${\tilde
  A}^{-0}$. Then $2\theta$ is equal to $\Phi + {\bar\Phi}$ or $\Phi -
{\bar\Phi}$, depending on the relative orientation of the triangles.
The minimum of $\cos 2\theta$ is obviously obtained when the two
triangles lie on two opposite sides of $A^{+0}$, corresponding to
$2\theta = \Phi + {\bar\Phi}$. {}From Fig.~\ref{isotriangles},
$\cos\Phi$ and $\cos{\bar\Phi}$ can be expressed in terms of
measurable quantities as
\bea
\cos\Phi & = & { {1\over 2} B^{+-} (1 + \adirpm) + B^{+0} - B^{00} (1
  + \adir00) \over \sqrt{2} \sqrt{B^{+-} (1 + \adirpm)} \sqrt{B^{+0}} } ~,
\nn\\
\cos\bar{\Phi} & = & { {1\over 2} B^{+-} (1 - \adirpm) + B^{+0} -
  B^{00} (1 - \adir00) \over \sqrt{2} \sqrt{B^{+-} (1 - \adirpm)}
  \sqrt{B^{+0}} } ~,
\eea
where $\adir00$ is defined analogously to $\adirpm$ in
Eq.~(\ref{observables}). By minimizing $\cos(\Phi + \bar\Phi)$ with
respect to $\adir00$, one finds that the minimum is obtained for
\beq
(\adir00)_{min} = \frac{\adirpm}{2}\frac{B^{+-}({1\over 2} B^{+-} -
B^{+0} - B^{00})}{B^{00}({1\over 2} B^{+-} + B^{+0} - B^{00})}~.
\eeq
The value of $\cos(\Phi + \bar\Phi)$ at the minimum is given by the
right-hand-side of Eq.~(\ref{superbound}).

Above we have derived a new upper bound on $|2\theta|$. This then
raises the question: is it possible to find a lower bound on this
quantity?  Unfortunately, the answer is no. This can be seen quite
clearly in Fig.~\ref{isotriangles}. Suppose that the two-triangle
isospin construction can be made for some nonzero value of $2\theta$.
It is then straightforward to show that one can always rotate $A^{+-}$
and ${\tilde A}^{+-}$ continuously around $W$ towards one another,
without changing $B^{00}$, until they lie on one line corresponding to
$\theta = 0$.  Thus, without measuring separately $\bd\to\pi^0\pi^0$
and $\bdbar\to\pi^0\pi^0$, one cannot put a lower bound on the penguin
pollution parameter.

\begin{figure}
\centerline{\epsfxsize 5.0 truein \epsfbox {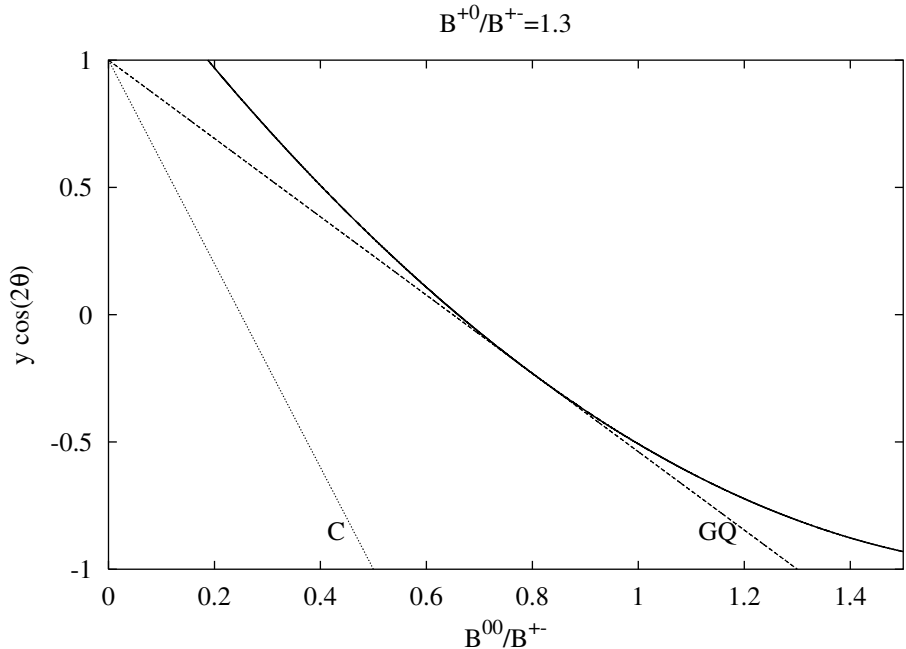}}
\caption{Bounds on $y\cos 2\theta$ as a function of $B^{00}/B^{+-}$,
  for $B^{+0}/B^{+-} = 1.3$. The curves represent the new bound [solid
  line, Eq.~(\protect\ref{superbound})], the Grossman-Quinn bound
  [dashed line, label GQ, Eq.~(\protect\ref{GQbound})] and the Charles
  bound [dotted line, label C, Eq.~(\protect\ref{Charlesbound})]. In
  all cases, the area below the curve is ruled out.}
\label{boundplot}
\end{figure}

We now turn to a comparison of the new bound on $\cos 2\theta$ with
the Grossman-Quinn and Charles bounds. The present world averages for
the $B\to\pi\pi$ branching ratios are (in units of $10^{-6}$)
\cite{pipi}:
\bea
BR(B \to \pi^+ \pi^-) & = & 4.4 \pm 0.9 ~,\nn\\
BR(B \to \pi^+ \pi^0) & = & 5.6 \pm 1.5 ~,\nn\\
BR(B \to \pi^0 \pi^0) & < & 5.7 ~(90\% ~{\rm C.L.}) ~.
\eea
For the purpose of illustration, we take central values,
$B^{+0}/B^{+-} = 1.3$, and compare the three bounds in
Fig.~\ref{boundplot}. In this figure, we plot $y\cos 2\theta$ as a
function of $B^{00}/B^{+-}$. In all cases, the region of parameter
space below the curve is ruled out. As expected, the new bound is
(almost) always more stringent than the Grossman-Quinn and Charles
bounds. (The new bound is equivalent to the Grossman-Quinn bound when
$2 B^{+0}/B^{+-} - 2 B^{00}/B^{+-} = 1$ [see Eq.~(\ref{SBGQequiv})],
i.e.\ for $B^{00}/B^{+-} = 0.8$.) Note also that the curves represent
the weakest possible lower bound on $\cos 2\theta$, obtained for
$y=1$.  Should $\adirpm$ be measured to be nonzero (i.e.\ $y<1$), this
will place a correspondingly stronger lower bound on $\cos 2\theta$.

\begin{figure}
\centerline{\epsfxsize 5.0 truein \epsfbox {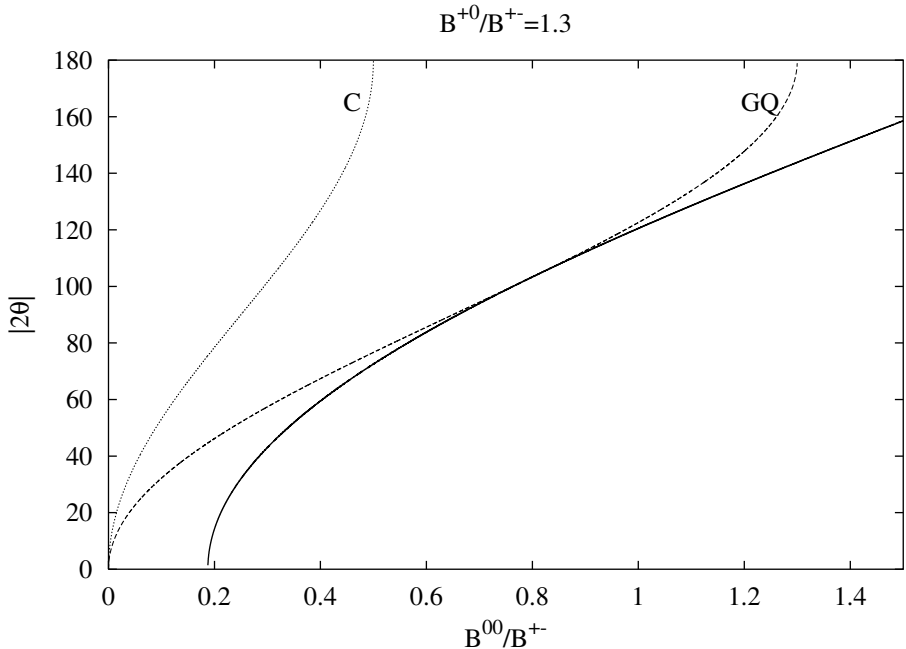}}
\caption{Constraints on $|2\theta|$ as a function of $B^{00}/B^{+-}$,
  for $B^{+0}/B^{+-} = 1.3$ and $y=1$. The curves represent the new
  bound [solid line, Eq.~(\protect\ref{superbound})], the
  Grossman-Quinn bound [dashed line, label GQ,
  Eq.~(\protect\ref{GQbound})] and the Charles bound [dotted line,
  label C, Eq.~(\protect\ref{Charlesbound})].  In all cases, values of
  $|2\theta|$ above the curve are ruled out.}
\label{boundplot2}
\end{figure}

The lower bound on $\cos 2\theta$ can be straightforwardly converted
into an upper bound on $|2\theta|$. This is shown explicitly in
Fig.~\ref{boundplot2}, where we plot the three bounds on $|2\theta|$
as a function of $B^{00}/B^{+-}$, for $B^{+0}/B^{+-} = 1.3$ and $y=1$.
In all cases, values of $|2\theta|$ above the curves are excluded.

One interesting feature of Fig.~\ref{boundplot} is that the lower
bound on $y\cos 2\theta$ seems to exceed unity for sufficiently small
values of $B^{00}/B^{+-}$. This implies that there is a lower limit on
$B^{00}/B^{+-}$, which one cannot see with the Grossman-Quinn or
Charles bounds. (This is also seen quite clearly in
Fig.~\ref{boundplot2}.) This lower limit, as well as an upper limit on
the same quantity, follows directly from the closure of the two
isospin triangles, which can be shown to imply that
\beq
{1\over 2} + {B^{+0} \over B^{+-}} - \sqrt{ {B^{+0} \over B^{+-}} (1 +
  y) } \le {B^{00}\over B^{+-}} \le {1\over 2} + {B^{+0} \over B^{+-}}
+ \sqrt{ {B^{+0} \over B^{+-}} (1 + y) } ~.
\label{B00bounds}
\eeq
(Equivalently, this constraint can be obtained from
Eq.~(\ref{superbound}) using the condition that $\cos 2\theta \le 1$.)
The limits are weakest for $y=1$. For $B^{+0}/B^{+-} = 1.3$, one finds
$0.19 \le B^{00}/B^{+-} \le 3.4$. An obvious implication of the
closure of the two isospin triangles is that a substantial deviation
of $2B^{+0}/B^{+-}$ from one, as demonstrated by the present central
values of $B^{+-}$ and $B^{+0}$, would be evidence for a sizeable
value of $B^{00}$.

This lower limit on $B^{00}/B^{+-}$ is useful for two reasons. First,
it will give experimentalists some knowledge of the branching ratios
for $\bd/\bdbar \to \pi^0\pi^0$. This in turn will help to anticipate
the feasibility of the full isospin analysis. Second, since the bound
on $B^{00}/B^{+-}$ relies only on the closure of the two triangles, it
will hold even in the presence of isospin-violating
electroweak-penguin contributions. On the other hand, it has been
pointed out by Gardner \cite{Gardner} that the triangles will not
close in the presence of other isospin-violating effects such as
$\pi^0$--$\eta,\eta'$ mixing.  Thus, the comparison of the actual
branching ratio $B^{00}$ with this bound may give some information
about the size of such isospin-violating effects.

Fig.~\ref{boundplot2} also shows that the upper bound on $|2\theta|$
deteriorates rather quickly as $B^{00}/B^{+-}$ increases above its
minimum value [Eq.~(\ref{B00bounds})]. This behaviour can be easily
understood: writing $B^{00}/B^{+-} = (B^{00}/B^{+-})_{min} + \Delta
B$, and expanding $\cos 2\theta$ around $2\theta = 0$,
Eq.~(\ref{superbound}) gives, for $y=1$,
\beq
(2\theta)^2 \le {4\sqrt{2} \over \sqrt{B^{+0}/B^{+-}}} \, \Delta B
- {2 \over B^{+0}/B^{+-}} \, \Delta B^2~.
\eeq
For $B^{+0}/B^{+-}$ of order unity, the coefficient of the linear term
is quite large, which causes the rapid deterioration of the upper
bound on $|2\theta|$ as $\Delta B$ increases. The only way to avoid
this is if $B^{+0}/B^{+-}$ is very large. However, since this
possibility is disfavoured experimentally, we conclude that the bound
on $|2\theta|$ will be very weak unless $B^{00}/B^{+-}$ happens to be
near its minimum allowed value. Of course, if $B^{00}/B^{+-}$ is above
its minimum, then it may well be possible to carry out the full
isospin analysis.

Although no lower limit can be obtained on the penguin-pollution angle
$|2\theta|$, we note that a lower bound can be derived for the
magnitude of the penguin amplitude $P$ from measurements of
$\bd(t)\to\pi^+\pi^-$ alone.  Consider again the expression for $|P|$
given in Eq.~(\ref{P+-}). This can be minimized with respect to
$\theta$, yielding
\beq
\tan\theta \vert_{minimum} = - \cot \alphaeff \left( {1-y \over 1+y}
\right) ~.
\eeq
The minimal value of $|P|^2$ is then
\beq
|P|^2_{min} = {B^{+-} (1 - y^2) \over 4 (1 - y \cos 2\alphaeff)} ~,
\eeq
where $\alphaeff$ is measured in the time-dependent rate of $B^0(t)\to
\pi^+\pi^-$ [Eq.~(\ref{B0(t)})].

To sum up, in the presence of penguin contributions, the CP asymmetry
in $\bd(t)\to\pi^+\pi^-$ no longer cleanly measures $\alpha$. It is
possible to remove this penguin ``pollution'' with the help of an
isospin analysis using all three $B\to\pi\pi$ decays and their charge
conjugates. However, this analysis requires separate measurements of
$BR(\bd \to \pi^0\pi^0)$ and $BR(\bdbar \to \pi^0\pi^0)$.  It will not
be possible to carry out the complete isospin analysis if only the sum
of this branching ratios can be measured. Nevertheless, we have shown
that an upper bound on the penguin-pollution angle $|2\theta|$ can be
obtained. This new bound is an improvement over bounds suggested
earlier by Grossman and Quinn, and by Charles. Indeed, since this
bound follows from the requirements that the two isospin triangles
close and have a common base, it is the most stringent bound possible
on $|2\theta|$. We have also derived a lower bound on the ratio of
branching ratios $B^{00}/B^{+-}$. If $B^{+0}/B^{+-}$ is larger than
unity, as is suggested by present central values, then $B^{00}/B^{+-}$
cannot be tiny and it may be possible to carry out the full isospin
analysis.  Finally, we have also shown how to obtain a lower bound on
the magnitude of the penguin amplitude $P$.

\section*{\bf Acknowledgments}

N.S. and R.S. thank D.L. for the hospitality of the Universit\'e de
Montr\'eal, where some of this work was done. We are grateful to Y.
Grossman for useful discussions. This work was partially supported by
the Israel Science Foundation founded by the Israel Academy of
Sciences and Humanities, by the U. S. -- Israel Binational Science
Foundation through Grant No.\ 98-00237, by NSERC of Canada, and by the
Department of Science and Technology, India.


\begin{thebibliography}{99}

\bibitem{CPreview} For a review, see, for example, {\it The BaBar
    Physics Book}, eds.\ P.F. Harrison and H.R. Quinn, SLAC Report
  504, October 1998.

\bibitem{betameas} T. Affolder {\it et al.} [CDF Collab.],
  \newprd{61}{2000}{072005}; A. Abashian {\it et al.} [BELLE Collab.],
  \prl{86}{2001}{2509}; B. Aubert {\it et al.} [BaBar Collab.],
  \prl{86}{2001}{2515}.

\bibitem{Dalitz} A.E. Snyder and H.R. Quinn, \prd{48}{93}{2139}; H.R.
  Quinn and J.P. Silva, \newprd{62}{2000}{054002}

\bibitem{BKK} A. Datta and D. London, hep-ph/0105073.

\bibitem{penguins} D. London and R. Peccei, \plb{223}{1989}{257}; M.
  Gronau, \prl{63}{1989}{1451}, \plb{300}{1993}{163}; B. Grinstein,
  \plb{229}{1989}{280}.

\bibitem{isospin} M. Gronau and D. London, \prl{65}{1990}{3381}.

\bibitem{GQ} Y. Grossman and H.R. Quinn, \newprd{58}{1998}{017504}.

\bibitem{Charles} J. Charles, \newprd{59}{1999}{054007}.

\bibitem{GPY} M. Gronau, D. Pirjol and T.M. Yan, \newprd{60}{1999}{034021};
A. J. Buras and R. Fleischer, \epjc{11}{1999}{93}.

\bibitem{Gardner} S. Gardner, \newprd{59}{1999}{077502}.

\bibitem{pipi} D. Cronin-Hennessy {\it et al.} [CLEO Collab.],
  \prl{85}{2000}{515}; D. Asner {\it et al.} [CLEO Collab.],
  hep-ex/0103040, submitted to {\it Phys. Rev. Lett.}; K. Abe {\it et
    al.} [Belle Collab.], hep-ex/0104030, submitted to {\it Phys. Rev.
    Lett.}; B. Aubert {\it et al.} [BaBar Collab.], hep-ex/0105061,
  submitted to {\it Phys. Rev. Lett.}.

\end{thebibliography}
\end{document}